\documentclass[12pt]{article}
\usepackage{amsmath,bbold,amsfonts,amssymb,graphics,xcolor}
\usepackage[T1]{fontenc}
\usepackage{graphicx}
\usepackage{tikz}

\makeatletter

\providecommand{\tabularnewline}{\\}

\def\TTbar{[T\bar{T}]}
\def\Tbar{\bar{T}}
\def\uqsu2{\mathcal{U}_{\mathsf{q}}(\mathfrak{su}_2)}
\newcommand{\q}{\mathsf{q}}

\def\th{\theta}

\def\half{{1\over{2}}}
\def\pd{\partial}

\newcommand{\beq}{\begin{equation}}
\newcommand{\eeq}{\end{equation}}
\newcommand{\bea}{\begin{eqnarray}}
\newcommand{\eea}{\end{eqnarray}}
\newcommand{\beaq}{\begin{eqnarray}}
\newcommand{\eeaq}{\end{eqnarray}}
\begin{document}
\begin{flushright}\footnotesize
\texttt{HU-EP-24/23}\\
\end{flushright}
\vspace{0.7cm}
\centerline{\Large \bf New Integrable RG flows with Parafermions}
\vskip 1cm
\centerline{\large Changrim Ahn\footnote{ahn@ewha.ac.kr} and Zoltan Bajnok\footnote{bajnok.zoltan@wigner.hu} }
\vskip 1cm
\centerline{\it$^{1}$Department of Physics, Ewha Womans University}
\centerline{\it Seoul 120-750, Korea}
\centerline{and}
\centerline{\it Institut f\"{u}r Physik, Humboldt-Universit\"{a}t zu Berlin} 
\centerline{\it Zum Gro\ss en Windkanal 2, 12489 Berlin, Germany}
\vskip 0.5cm
\centerline{\it$^{2}$HUN-REN Wigner Research Centre for Physics,}
\centerline{\it  Konkoly-Thege Miklos ut 29-33, 1121  Budapest, Hungary}
\vskip 0.8cm
\centerline{\small PACS: 11.25.Hf, 11.55.Ds}
\vskip 0.8cm
\centerline{\bf Abstract}

We consider irrelevant deformations of massless RSOS scattering theories by an infinite number of higher $\TTbar_{s+1}$ operators which introduce extra non-trivial CDD factors between left-movers and right-movers.
It is shown that the resulting theories can be UV complete after bypassing typical Hagedorn-like singularities if the coefficients of the deformations are fine-tuned.
In this way, we have discovered that only two new UV complete QFTs are associated with a $\mathcal{M}_p\ (p=3,4,\cdots)$ minimal CFT based on the integrable structure of the RSOS scattering theory. 
One is the massless $\mathbb{Z}_{p-1}$ parafermionic sinh-Gordon models (PShG) with a self-dual coupling constant. This correspondence is confirmed by showing that the scale-dependent vacuum energies computed by the thermodynamic Bethe ansatz  based on the $S$-matrices match those from the quantization conditions for the PShG models using the reflection amplitudes. 
The other UV QFT is reached from $\mathcal{M}_p$ by following the roaming trajectory of the $\mathbb{Z}_{p-1}$ parafermionic minimal series.

\newpage

\section{Introduction}

Integrable quantum field theories (QFTs) have special properties that can provide quantitative 
methods to study QFTs  non-perturbatively \cite{sasha}. 
One interesting and relevant example is finding non-trivial renormalization group (RG) flows that exhibit fixed points in the IR (infra-red) limit \cite{zamolRG}. 
In general, this problem is very difficult to solve analytically since asymptotically free QFTs become non-perturbative at the IR  scale.
Fortunately, integrable QFTs in two dimensions have provided quantitative methods to address  this problem. 
A basic framework is to use the exact $S$-matrices, defined in the IR theories at infinite volume, to construct the thermodynamic Bethe ansatz (TBA) equations to generate scaling functions at all scales from the UV down to the IR  \cite{alyosha_tba}. 

To find such CFT pairs connected by the RG, a traditional approach is to ask ``Which relevant operator can deform the UV CFT to maintain integrability and what is the corresponding IR CFT and its irrelevant perturbations?'' 
Once the perturbation is integrable and the scattering matrix is known, the TBA method can be applied to describe exact flows on either scale.
 In this way, exact RG flows have been proposed by either $S$-matrices or conjecturing the TBAs \cite{alyosha_tba_flow, alyosha_tba_cosetflow, zamzam_flow, FSZ, DDT, AKRZ}. A main technical difficulty arises from the fact that  integrability is preserved typically by deformations of a single relevant field.\footnote{In higher rank theories there could be multi-parameter integrable deformations.} 
 This would mean that there can be more pairs of CFTs connected by such exact computations. 
 In particular, it is conceivable that the same IR CFT can be reached from several different UV CFTs along with well-chosen relevant operators. 
 
 This possibility poses an important question,
 ``Can we completely classify UV CFTs that can be reached from a given IR CFT?''
 To answer this question, we need to classify all possible integrable deformations of the IR CFT by irrelevant operators.
This problem has been studied recently in \cite{AhnLeC} by exploiting the recent developments claiming that a special class of irrelevant operators belonging to the energy-momentum tensor operators and their descendants can preserve integrability \cite{SZ,CNST}. 
 Constrained to diagonal scattering theories, it has been found that several UV CFTs can be reached from certain minimal IR CFTs. 
 For example, there are about 4 UV CFTs which all flow into the critical Ising CFT. 
 Some of these are already noticed in the previous works and some are new. 
 For those new flows, the TBA equations can provide only the central charges of the UV CFTs, often not enough to identify them.
 
 In this paper, we focus on non-diagonal scattering theories for which this approach can be more restrictive. In particular, we find the UV CFTs and their relevant perturbation, such that the integrable flow ends at the unitary minimal CFTs, the most studied CFTs.

\section{Massless kink $S$-matrices and their TBA}

\subsection{Restricted sine-Gordon model and Massive kinks}

We start with a minimal CFT ${\mathcal M}_p$ with a central charge $c=1-6/(p(p+1))$ perturbed by the least relevant operator $\Phi_{\rm pert}$ with a conformal dimension $h=(p-1)/(p+1)$,  whose formal action
can be written as
\bea
{\cal S}_{\lambda}={ \cal S}_{{\cal M}_p} + \lambda \int d^2x \  \Phi_{\rm pert}.
\label{pertmin}
\eea
This model is a well-known integrable QFT \cite{sasha}, which is related to the quantum sine-Gordon model with
$\uqsu2$ quantum group symmetry, where the deformation parameter $\q$ is related to $p$ \cite{BerLeC}.
The $S$-matrix of \eqref{pertmin} for a given $p$ is obtained by truncating the multi-soliton and antisoliton Hilbert space for $\q$ a root of unity.
On-shell particles are ``kinks'' connecting two adjacent vacua denoted by the corresponding spins
\bea
a\ \Bigg\uparrow\ b\ =\ K_{ab}(\theta),\qquad a,b=0,\half,\cdots,\frac{p}{2}-1,\quad {\rm with}\quad |a-b|=\half,
\eea
When $\lambda<0$, particles are massive and each particle carries an energy and momentum 
\bea
E=M\cosh\theta,\qquad P=M\sinh\theta,
\label{disper}
\eea
where the mass $M$ is related to $\lambda$ by \cite{Alyoshamassmu}
\bea
\pi|\lambda|=\frac{(p+1)^2}{(p-1)(2p-1)}
\left[\frac{\gamma\left(\frac{3p}{p+1}\right)}{\gamma\left(\frac{1}{p+1}\right)}\right]^{1/2}
\left[\frac{\sqrt{\pi}M\Gamma\left(\frac{p+1}{2}\right)}{2\Gamma\left(\frac{p}{2}\right)}\right]^{4/(p+1)}.
\eea
Here $\gamma(x)=\Gamma(x)/\Gamma(1-x)$. 
\begin{figure}
\label{figkinkS}
\begin{center}
\begin{picture}(100,100)(0,0)
\thicklines
\put(0,0){\line(1,1){80}}
\put(0,0){\vector(1,1){25}}
\put(40,40){\vector(1,1){25}}
\put(80,0){\line(-1,1){80}}
\put(80,0){\vector(-1,1){25}}
\put(40,40){\vector(-1,1){25}}
\put(0,40){$d$}
\put(40,0){$a$}
\put(80,40){$b$}
\put(40,80){$c$}
\put(-10,-10){$\theta_1$}
\put(80,-10){$\theta_2$}
\end{picture}
\end{center}
\caption{Kink $S$-matrix}
\end{figure}
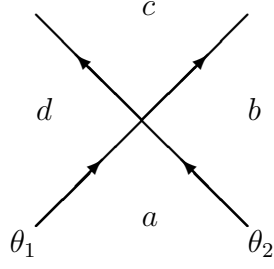
A two-particle $S$-matrix elements for a scattering process 
\bea
K_{da}(\theta_1) + K_{ab}(\theta_2)\rightarrow K_{dc}(\theta_2) + K_{cb}(\theta_1)
\eea
in Fig.1 are  given by\footnote{In a recent paper \cite{Shota}, a generalized crossing symmetry related to a non-invertible symmetry allows new $S$-matrix without the prefactor $\left(X_{db}^{ac}\right)^{\frac{i\theta}{2\pi}}$. This gauge factor does not change the TBA.}
\bea
\label{kinkS}
{S_{p}(\theta)}^{ab}_{dc}=U(\theta)\, \left(X_{db}^{ac}\right)^{\frac{i\theta}{2\pi}}
\left[
\left(X_{db}^{ac}\right)^{\frac{1}{2}}\sinh\left(\frac{\theta}{p}\right)\,
\delta_{db}  +\sinh\left(\frac{i\pi -\theta}{p}\right)\,
\delta_{ac}
\right], 
\eea
where $U(\theta)$ is a well-known scalar factor of the sine-Gordon model ($\theta=\theta_1-\theta_2$) given by 
\bea
U(\theta)=\frac{1}{\sinh\frac{1}{p}(\theta-i\pi)}\exp\,\left[\int_{-\infty}^{\infty}\frac{dk}{k}\,
\frac{\sinh\frac{k\pi(p-1)}{2}}{2\sinh\frac{k\pi p}{2}\,\cosh\frac{k\pi}{2}}e^{ik\theta}\right],
\eea
and 
\bea
X_{db}^{ac}=\frac{[2a+1][2c+1]}{[2d+1][2b+1]} 
\eea
with the $\q$-number defined by
\bea
[n] = { {\q^n - \q^{-n} }\over {\q-\q^{-1} } } , ~~~~{\rm with} ~~
\q=-\exp\left({-i\pi\over{p} }\right).
\eea
The exact two-particle $S$-matrices are given by a quantum group reduction of the scatterings of the quantum sine-Gordon model, 
where the multi-soliton and antisoliton Hilbert space is truncated depending on the values of the quantum group deformation parameter $\q$.

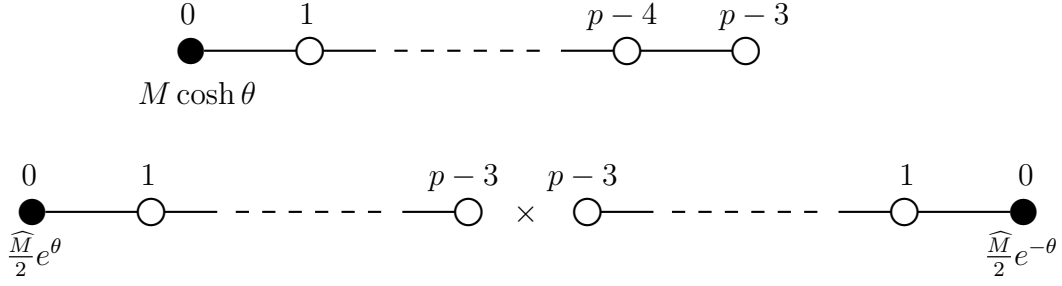
\begin{figure}
\begin{picture}(310,60)(0,-20)
\thicklines
\put(95,20){\circle*{10}}
\put(100,20){\line(1,0){35}}
\put(140,20){\circle{10}}
\put(145,20){\line(1,0){20}}
\multiput(172,20)(10,0){6}{\line(1,0){5}}
\put(305,20){\circle{10}}
\put(300,20){\line(-1,0){35}}
\put(260,20){\circle{10}}
\put(255,20){\line(-1,0){20}}
\put(91,30){$0$}
\put(136,30){$1$}
\put(295,30){$p-3$}
\put(245,30){$p-4$}
\put(75,0){$M\cosh\theta$}
\end{picture}
\begin{picture}(400,40)(-30,0)
\thicklines
\put(5,20){\circle*{10}}
\put(10,20){\line(1,0){35}}
\put(50,20){\circle{10}}
\put(55,20){\line(1,0){20}}
\multiput(82,20)(10,0){6}{\line(1,0){5}}
\put(170,20){\circle{10}}
\put(165,20){\line(-1,0){20}}
\put(215,20){\circle{10}}
\put(340,20){\line(1,0){35}}
\put(335,20){\circle{10}}
\put(220,20){\line(1,0){20}}
\multiput(247,20)(10,0){6}{\line(1,0){5}}
\put(330,20){\line(-1,0){20}}
\put(380,20){\circle*{10}}
\put(1,30){$0$}
\put(46,30){$1$}
\put(155,30){$p-3$}
\put(-5,0){$\frac{\widehat{M}}{2}e^{\theta}$}
\put(187,16){$\times$}
\put(378,30){$0$}
\put(333,30){$1$}
\put(200,30){$p-3$}
\put(365,0){$\frac{\widehat{M}}{2}e^{-\theta}$}
\end{picture}
\caption{(above) $A_{p-2}$ Dynkin diagram for massive and (below) conformal RSOS theories}
\label{AnDynkin}
\end{figure}

A fundamental tool to investigate scaling functions such as the vacuum energy for a given scale is TBA,  which minimizes the free energy that is expressed by densities of on-shell particle states subject to periodic boundary condition (PBC) \cite{alyosha_tba}.
For non-diagonal scattering theories such as for the RSOS scattering theory above, deriving the TBA involves the difficult step of diagonalizing an inhomogeneous transfer matrix. 
This problem has been well studied both in QFTs and statistical lattice models in two dimensions. 
The eigenvalues are expressed in terms of new degrees of freedom, the so called ``magnons'' of $p-3$ different species, as well as the ``physical'' on-shell particles
\bea
\mathbb{T}_{\rm RSOS}(\{\theta_i\})\vert\Psi\rangle=
\Lambda\left(\{\theta_i\},\oplus_{a=1}^{p-3}\{\lambda^{(a)}_j\}\right)\vert\Psi\rangle.
\label{RSOSTransf}
\eea
The TBA equations are obtained for the pseudo-energies corresponding to the densities of magnons and particles,
\bea
\epsilon_a(\theta)=\delta_{a0}MR\cosh\theta-\sum_{b=0}^{p-3}\mathbb{I}_{ab}
\varphi\star\ln\left(1+e^{-\epsilon_b}\right)(\theta),\quad a=0,\cdots,p-3,
\eea
where the ``universal'' kernel $\varphi$ is given by
\bea
\varphi(\theta)=\frac{1}{\cosh\theta},
\eea
 $\star$ is the convolution and $\mathbb{I}$ is the adjacency matrix of the $A_{p-2}$ Dynkin diagram, namely 
$\mathbb{I}_{ab}$ is $1$ if nodes $a$ and $b$ are connected in Fig.\ref{AnDynkin} (above) and zero otherwise.

\subsection{Massless kinks scattering}

We can think of the original conformal field theory as the  $\lambda \to 0^-$ limit of the massive scattering theory. In this limit the scattering states become massless with a dispersion relation $E^2=P^2$ between the energy ($E$) and momentum ($P$) carried by the asymptotic particles.
The $E=+P$ and $E=-P$ cases describe right-moving ($R$) and left-moving ($L$) massless states, respectively.
These states can be thought of as an extremely relativistic limit of massive states by rescaling all rapidities
$\theta_i\to \hat{\theta}_i+\Lambda$ for  ($R$) and 
$\theta_i\to \hat{\theta}_i-\Lambda$ for  ($L$) in \eqref{disper} and 
taking limits  $M\to 0,\ \Lambda\to\infty$ while keeping
$M e^{\Lambda}=\widehat{M}$ finite. 
We will take $\hat{\theta}$ as the rapidity from now on with which the dispersion  relation is expressed as
\bea
E=\pm  P= \frac{\widehat{M}}{2}\, e^{\pm\hat{\theta}},\qquad +=R,\quad -=L.
\label{RLpart}
\eea

In this limit, $S$-matrices between the same types ($R$ or $L$) of massless particles are the same as the massive ones \eqref{kinkS} because $\theta_1-\theta_2=\hat{\theta}_1-\hat{\theta}_2$
\bea
S^{RR}_{p}(\theta)=S^{LL}_{p}(\theta)=S_{p}(\theta).
\label{masslessRR}
\eea
Scatterings between $R$ and $L$ particles become independent of the rapidities because $|\theta_1-\theta_2|\to\infty$ and the kernels between these particles vanish in the TBA.
Therefore,  the TBA equations for massless kink theories are described by two separate sets ($R$ and $L$) of equations,\footnote{We will replace $\hat{\theta}$ with $\theta$ for the rapidity from now on and use superindex `$+$' for $R$-type and `$-$' for  $L$-type in the TBA equations.}
\bea
\epsilon^{\pm}_a(\theta)=\delta_{a0}\frac{\widehat{M}R}{2}e^{\pm\theta}-\sum_{b=0}^{p-3}\mathbb{I}_{ab}
\varphi\star\ln\left(1+e^{-\epsilon^{\pm}_b}\right)(\theta),\quad a=0,\cdots,p-3,
\label{CFTTBA}
\eea
where the adjacency matrix $\mathbb{I}_{ab}$ is given by Fig.\ref{AnDynkin} (below).
This TBA system describes the scale-invariant minimal CFT $\mathcal{M}_p$ since any change of the scale $R$ can be absorbed into a shift of the rapidity $\theta$.

Eq.\eqref{pertmin} with $\lambda>0$ was shown to generate an RG flow from a UV $\mathcal{M}_{p}$ CFT to the $\mathcal{M}_{p-1}$ IR CFT \cite{zamolRG}.  We can also think of the same flow the opposite direction, i.e. connecting the  IR $\mathcal{M}_{p-1}$ minimal model to the UV $\mathcal{M}_{p}$ minimal model.  As we already formulated the integrable scattering description of the $\mathcal{M}_{p}$ minimal model, we will concentrate on the (shifted) flow when the IR CFT is the $\mathcal{M}_{p}$ minimal model and the UV CFT is the $\mathcal{M}_{p+1}$ minimal model.  This is a conceptual change of point of view. We would like to describe the flow by deforming the IR CFT with irrelevant operators. In the scattering language it means to introduce non-trivial scatterings between the left and right moving particles.  The quantitative evaluation of the corresponding scaling function, such as the effective central charge, can be obtained by the conjectured TBA system in \cite{alyosha_tba_flow}.

\begin{figure}
\centerline{
\begin{tikzpicture}
\draw[thick] (0,0) -- (1.5,0) (1.9,0) -- (2.7,0) (5.1,0) -- (5.9,0) (6.3,0) -- (7.8,0) ;
\node at (7.8,-0.6) {$\frac{\widehat{M}}{2}e^{-\theta}$};
\node at (0,-0.6) {$\frac{\widehat{M}}{2}e^{\theta}$};
\node at (-0.1,0.5) {$0$};
\node at (1.7,0.5) {$1$};
\node at (6.1,0.5) {$p-3$};
\node at (7.9,0.5) {$p-2$};
\draw[thick,dashed] (2.9,0) -- (4.9,0);
\draw[thick] (1.7,0) circle (0.2);
\draw[thick] (6.1,0) circle (0.2);
\fill[red]  (-0.1,0) circle (0.2);
\fill[blue]  (7.9,0) circle (0.2);
\end{tikzpicture}
}
\caption{$A_{p-1}$ Dynkin diagram for massless RSOS theories}
\label{An1Dynkin}
\end{figure}
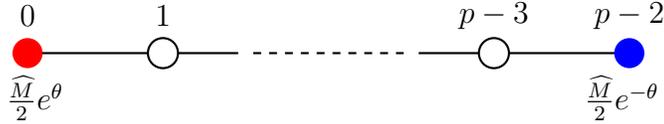

This conjecture was confirmed based on the massless scattering theory in \cite{FSZ}.
In this work, the $S^{(p)RL}_{\rm RSOS}(\theta)$ scatterings between the left- and right-moving kinks were determined by the  crossing-unitarity relations and the Yang-Baxter equation (YBE) betwen this and the
$S^{(p)RR}_{\rm RSOS},\ S^{(p)LL}_{\rm RSOS}$ scattering matrices giving 
\bea
\label{masslessRL}
S^{RL}_{p}(\theta)=\frac{\tilde{U}(\theta)}{U(\theta+\frac{ip\pi}{2})}\, S_{p}\left(\theta+\frac{ip\pi}{2}\right),\quad
S^{LR}_{p}(\theta)=-\frac{\tilde{U}(\theta)}{U(\theta-\frac{ip\pi}{2})}\, S_{p}\left(\theta-\frac{ip\pi}{2}\right),
\eea
where $\tilde{U}(\theta)$ is given by \cite{FSZ}
\bea
\tilde{U}(\theta)=\frac{1}{\cosh\frac{1}{p}(\theta-i\pi)}\exp\,\left[-\int_{-\infty}^{\infty}\frac{dk}{k}\,
\frac{\sinh\frac{k\pi}{2}}{2\sinh\frac{k\pi p}{2}\,\cosh\frac{k\pi}{2}}e^{ik\theta}\right].
\eea
Introducing non-trivial scatterings between left and right movers modifies the finite volume spectrum. As the RSOS part of the left-right scattering is shifted we have to diagonalize an inhomogenous transfer matrix, in which the inhomogeneities corresponding to the right movers are shifted. This shift implies that the right movers couple to the  $p-3$ magnons differently: we have to flip that part of the Dynkin diagram. This was rigorously shown in \cite{FSZ}  for $p=4$ and the authors argued that  the same happens with higher $p$'s. This leads to the conjectured TBA by Zamolodchikov, whose the adjacency matrix $\mathbb{I}_{ab}$ is given by the  Dynkin diagram in Fig.\ref{An1Dynkin},
\bea
\epsilon_a(\theta)=\delta_{a0}\frac{\widehat{M}R}{2}e^{\theta}
+\delta_{a,p-2}\frac{\widehat{M}R}{2}e^{-\theta}-\sum_{b=0}^{p-2}\mathbb{I}_{ab}
\varphi\star\ln\left(1+e^{-\epsilon_b}\right)(\theta),\quad a=0,\cdots,p-2.
\label{massflowTBA}
\eea
The ground-state energy at the scale $R$ and effective central charge are defined by
\bea
E_0(R)=-\frac{\widehat{M}}{4\pi}\int_{-\infty}^{\infty}\left[e^{\theta}\ln\left(1+e^{-\epsilon_0(\theta)}\right)+
e^{-\theta}\ln\left(1+e^{-\epsilon_{p-2}(\theta)}\right)\right]d\theta
=-\frac{\pi C_{\rm eff}(\widehat{M}R)}{6R}.
\eea
This TBA system generates a RG flow from the  $\mathcal{M}_{p+1}$ CFT ($R= 0$) to the $\mathcal{M}_{p}$ CFT ($R=\infty$).

In fact, we should interpret this set of $S$-matrices 
$S^{RR}_{p}(\theta),S^{LL}_{p}(\theta),S^{RL}_{p}(\theta),S^{LR}_{p}(\theta)$ which describes the CFT $\mathcal{M}_{p}$ deformed by an irrelevant operator $\Phi_{3,1}$ since the $S$-matrix should be defined in the infinite volume or the IR limit.
The TBA given in \eqref{massflowTBA} interpolates through an RG flow from the IR CFT to the UV CFT  
$\mathcal{M}_{p+1}$.

Certainly, different irrelevant deformations will generate different flows to other UV CFTs perturbed by some relevant operators. 
For example, one simple possibility is to choose
\bea
\label{masslessRL2}
S^{RR}_{p}(\theta)=S^{LL}_{p}(\theta)=S^{RL}_{p}(\theta)=S^{LR}_{p}(-\theta)=S_p(\theta).
\eea

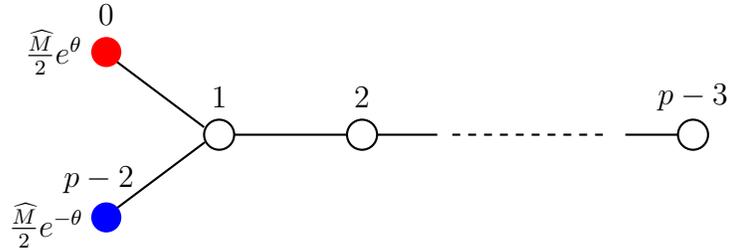
\begin{figure}
\centerline{
\begin{tikzpicture}
\draw[thick] (0.3,1) -- (1.5,0.1) ;
\draw[thick] (0.31,-1) -- (1.51,-0.1) ;
\draw[thick] (1.9,0) -- (3.4,0) (3.8,0) -- (4.6,0) (7.1,0) -- (7.8,0) ;
\node at (-0.5,1.1) {$\frac{\widehat{M}}{2}e^{\theta}$};
\node at (-0.6,-1.2) {$\frac{\widehat{M}}{2}e^{-\theta}$};
\node at (0.2,1.6) {$0$};
\node at (1.7,0.5) {$1$};
\node at (3.6,0.5) {$2$};
\node at (8,0.5) {$p-3$};
\node at (0.1,-0.6) {$p-2$};
\draw[thick,dashed] (4.8,0) -- (6.8,0);
\draw[thick] (1.7,0) circle (0.2);
\draw[thick] (3.6,0) circle (0.2);
\draw[thick] (8,0) circle (0.2);
\fill[blue]  (0.2,-1.1) circle (0.2);
\fill[red]  (0.2,1.1) circle (0.2);
\end{tikzpicture}
}
\caption{$D_{p-1}$ Dynkin diagram for massless RSOS theories}
\label{D_type}
\end{figure}
This $RL$ $S$-matrices do not have shifts in the rapidity shown in \eqref{masslessRL}.
In this case, a $R$-particle will scatter with both $R$- and $L$-particles with the same $S$-matrix in the virtual process when  we move it around the periodic volume.
The resulting PBC equation is obtained by the same transfer matrix as \eqref{RSOSTransf}. 
In the same way, a $L$-particle will generate the same transfer matrix after scattering with all other particles.
This common transfer matrix can be diagonalized and its eigenvalue is expressed in terms of the magnons.
The resulting TBA equations are still given by \eqref{massflowTBA} but with a different adjacency matrix in Fig.\ref{D_type} where both $R$- and $L$-nodes are connected to the first magnon node.
In fact this TBA has been already conjectured for a Parafermion CFT perturbed by a relevant bilinear parafermionic fields in \cite{Fateev} without specifying the $S$-matrices. The massless $S$-matrices in \eqref{masslessRL2} provide them.
The two cases above show clearly how the same $LL$ and $RR$ but different $RL$ $S$-matrices based on the same IR CFT can describe different QFTs that reach at distinct CFTs in the UV limit.

\section{$\TTbar$ deformations of RSOS $S$-matrices}
In this section, we deform an integrable QFT with a special set of irrelevant operators that maintain integrability with a deformed $S$-matrix. 
When a CFT is perturbed by a relevant operator $\Phi_{\rm pert}$, the holomorphic and anti-holomorphic conserved charges rearrange to satisfy new
conserved local currents satisfying the continuity equations
\beq
\partial_{\bar{z}} T_{s+1} = \partial_{z}  \Theta_{s-1}, ~~~~~~
\partial_{z} \Tbar_{s+1} = \partial_{\bar{z}} \bar{\Theta}_{s-1},
\eeq
where $s$ is a positive integer 
with $s+1$ and $-(s+1)$  the spins of $T_{s+1}$ and $\Tbar_{s+1}$ respectively.  
For $s=1$ these are the components of the universal stress-energy tensor and the conserved charges are the left and right components of the energy and momentum.   
For higher $s$,  the above currents depend on the model,  in particular the choice of $\Phi_{\rm pert}$.  
Smirnov and Zamolodchikov showed that from these one can construct well-defined local operators $\TTbar_s$:
\beq
\label{Xsdef}
\TTbar_s  = T_{s+1} \Tbar_{s+1} - \Theta_{s-1} \bar{\Theta}_{s-1}
\eeq
with scaling dimension $({\rm mass})^{2(s+1)}$.    
More importantly, perturbation by such operators preserves integrability \cite{SZ}.       
Thus,  we can consider the theory defined by the action 
\bea
{\cal S}_{\lambda}={\cal S}_{\mathcal{M}_{p}}+\lambda\int d^2x \ \Phi_{\rm 3,1}
+ \sum_{s\geq 1}  \alpha_s \, \int d^2 x \,  \TTbar_s 
\label{pertminTT}
\eea
where $\alpha_s$ are coupling constants of scaling dimension $[{\rm mass}]^{-2s}$.


\subsection{CDD factors}

It is known that the perturbation by the irrelevant operators $\alpha_s\TTbar_s$ simply modifies the original 
S-matrix by a CDD factor \cite{SZ},
\bea
S^{(s)}_{\rm CDD}(\theta)=e^{i g_s\sinh s\theta},
\eea
where a dimensionless  coefficient $g_s$ is given by
\bea
\label{massivegs}
 g_s=-\alpha_s M^{2s}.
\eea
Being CDD, this additional $S$-matrix factor acts as a scalar factor which does not change the matrix structure.
If the particles become massless, this relation changes by the massless scaling limit $\theta\to \hat{\theta}\pm\Lambda, \Lambda\to\infty, M\to 0$ with $M e^{\Lambda}=\widehat{M}$ finite as explained in \eqref{RLpart}.
Then, the CDD $S$-matrices become
\bea
S^{(s)RR}_{\rm CDD}=S^{(s)LL}_{\rm CDD}=1,\qquad  S^{(s)RL}_{\rm CDD}(\theta)=\exp[i \hat{g}_s e^{s\theta}],\qquad  S^{(s)LR}_{\rm CDD}(\theta)=\exp[i \hat{g}_s e^{-s\theta}].
\eea
Here we replaced $\hat{\theta}$ with $\theta$ and defined
\bea
\label{masslessgs}
\hat{g}_s=-\alpha_s \widehat{M}^{2s}.
\eea

If we add these irrelevant operators to the deformed minimal model with $\Phi_{\rm pert}$ as in \eqref{pertminTT}, the diagonal CDD factor $S$-matrix becomes
\bea
\label{CDDTT}
S_{\rm CDD}^{(\alpha)RL}(\theta)=\prod_{s=1}^{\infty}\exp[i \hat{g}_s e^{s\theta}],
\eea
which will modify only the $RL$ and $LR$ scattering processes.
We use a super-index $(\alpha)$ to denote the set of coefficients $\alpha_s$ in \eqref{pertminTT}.
The resulting $S$-matrices that replace those in \eqref{masslessRR} and \eqref{masslessRL} are
\bea
S^{(\alpha)RR}_{p}(\theta)&=&S^{(\alpha)LL}_{p}(\theta)=S_{p}(\theta),
\label{masslessRRTT}\\
S^{(\alpha)RL}_{p}(\theta)&=&S^{(\alpha)LR}_{p}(-\theta)=S_{\rm CDD}^{(\alpha)RL}(\theta)\cdot S_{p}(\theta+\frac{i p \pi }{2}).
\label{masslessRLTT}
\eea

The theory \eqref{pertminTT} is described by these new RSOS massless $S$-matrices and a corresponding TBA equation.
Since the CDD factors are scalar functions, the magnonic structure should be the same as the TBA in \eqref{massflowTBA} with an additional link between the two colored nodes in Fig.3 by a kernel given by
\bea
\varphi_{RL}(\theta)=\varphi_{LR}(-\theta)=\frac{1}{i}\frac{\partial}{\partial\theta}\ln S^{(\alpha)RL}_{\rm CDD}(\theta).
\eea 
These kernels will depend on all possible sets of the parameters $\alpha_s$.
We will choose such sets which will generate RG flows into some well-defined UV CFTs.
For this, we can consider a much simplified version of the TBA, namely, a plateaux equation as used in \cite{AhnLeC}.

\subsection{TBA and UV complete theories}

If the theory is UV complete, the TBA should have well-defined solutions in $r\equiv\widehat{M}R\to 0$ limit where
the pseudo-energies have constant values  in a wide region centered at $\theta=0$. 
Then, the TBA equations are reduced to much simpler algebraic equations between these plateaux values.\footnote{If the UV theory is an irrational CFT, these plateaux may not appear. Even so, it turns out that these equations still give correct UV central charges.}

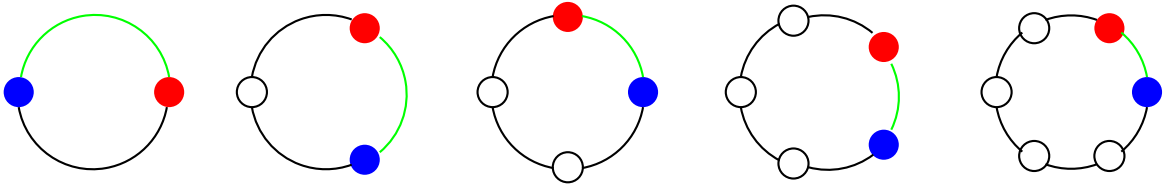
\begin{figure}
\begin{tikzpicture}
\fill[red]  (2,2) circle (0.2cm);
\fill[blue]  (0,2) circle (0.2cm);
\draw[green,thick] (2,2.2) arc (10:170:1cm);
\draw[thick] (0,1.8) arc (190:350:1cm);
\draw[thick] (3.1,2) circle (0.2cm);
\fill[red]  (4.6,2.85) circle (0.2cm);
\fill[blue]  (4.6,1.1) circle (0.2cm);
\draw[thick] (3.1,2.2) arc (170:70:1cm);
\draw[thick] (3.1,1.8) arc (190:290:1cm);
\draw[green,thick] (4.8,1.2) arc (-50:50:1cm);
\draw[thick] (6.3,2) circle (0.2cm);
\draw[thick] (7.3,1) circle (0.2cm);
\fill[red]  (7.3,3) circle (0.2cm);
\fill[blue]  (8.3,2) circle (0.2cm);
\draw[thick] (6.3,2.2) arc (170:100:1cm);
\draw[thick] (6.3,1.8) arc (190:260:1cm);
\draw[thick] (8.3,1.8) arc (-10:-80:1cm);
\draw[green,thick] (8.3,2.2) arc (10:80:1cm);
\draw[thick] (9.6,2) circle (0.2cm);
\draw[thick] (10.3,2.95) circle (0.2cm);
\draw[thick] (10.3,1.05) circle (0.2cm);
\fill[red]  (11.5,2.6) circle (0.2cm);
\fill[blue] (11.5,1.3)  circle (0.2cm);
\draw[thick] (9.6,2.2) arc (170:118:1cm);
\draw[thick] (9.6,1.8) arc (190:242:1cm);
\draw[thick] (10.5,3) arc (102:50:1cm);
\draw[thick] (10.5,1) arc (255:307:1cm);
\draw[green,thick] (11.6,1.5) arc (-26:26:1cm);
\draw[thick] (13,2) circle (0.2cm);
\draw[thick] (13.5,2.85) circle (0.2cm);
\draw[thick] (13.5,1.15) circle (0.2cm);
\draw[thick] (14.5,1.15) circle (0.2cm);
\fill[red]  (14.5,2.85) circle (0.2cm);
\fill[blue]  (15,2) circle (0.2cm);
\draw[thick] (13,2.2) arc (170:130:1cm);
\draw[thick] (13,1.8) arc (190:230:1cm);
\draw[green,thick] (15,2.2) arc (10:50:1cm);
\draw[thick] (15,1.8) arc (-10:-50:1cm);
\draw[thick] (13.65,1.04) arc (250:290:1cm);
\draw[thick] (13.65,2.96) arc (110:70:1cm);
\end{tikzpicture}
\caption{$\widehat{A}_{p-1}$ ($p=3,4,5,6,7$) affine Dynkin diagrams for new scattering theories with blue ($R$), red ($L$), and empty (magnons) nodes. The green links denote shifted universal kernels by $\alpha$.}
\label{An1DynkinTT}
\end{figure}

These equations are given by
\newcommand{\as}{\mathsf{a}}
\bea
x_n&=&(1+x_{n-1})^{1/2}(1+x_{n+1})^{1/2},\quad n=1,\cdots,p-3,\\
x_0&=&(1+x_{1})^{1/2}(1+x_{p-2})^{\as},\quad
x_{p-2}=(1+x_{p-3})^{1/2}(1+x_{0})^{\as},
\eea
where we have defined 
\bea
x_n\equiv e^{-\epsilon_n(0)},\quad 
\as=\int_{-\infty}^{\infty} \varphi_{RL}(\theta)\, \frac{d\theta}{2\pi}.
\eea
This set of algebraic equations can be easily solved either numerically or even analytically depending on the 
exponent $\as$. 
Since the pseudo energies are real, we need to find $\as$ which gives real solutions.
It can be easily checked that only $\as=0$ and $\as=1/2$ give real solutions.
The $\as=0$ case corresponds to the TBA \eqref{massflowTBA} because no CDD factors are added.
The $S$-matrices between $L$- and $R$-particles are given by \eqref{masslessRL}.
If $\as=1/2$, the TBA is described by the affine Dynkin diagram $\widehat{A}_{p-1}$ in Fig.\ref{An1DynkinTT}. 
The only kernel that comes from the CDD, which satisfies the crossing-unitarity should be the usual 
\bea
\varphi_{RL}(\theta)=\varphi_{LR}(-\theta)=\frac{1}{\cosh(\theta-\alpha)},
\label{shiftedkernel}
\eea
which comes from
\bea
S_{\rm CDD}^{RL}(\theta)=S_{\rm CDD}^{LR}(-\theta)=-\tanh\left(\frac{\theta-\alpha}{2}-\frac{i\pi}{4}\right).
\label{RLcdd}
\eea
Comparing with \eqref{CDDTT}, we can find that the coefficients of the $\TTbar_s$ should be
\bea
\hat{g}_s=-2\, i^s\, e^{-s\alpha},\quad {\rm with}\quad s={\rm odd}.
\eea

By solving the TBA numerically, we have found that the spectrum becomes complex when  the constant $\alpha$ complex. 
Therefore, we will take $\alpha$ as real from now on.
We first start with $\alpha=0$ and will present $\alpha\neq 0$ in the sect.6.

When $\alpha=0$, the $\varphi_{RL}$ becomes the universal kernel
and  the new TBA is the same as \eqref{massflowTBA} with the additional link connecting the nodes $0$ and $p-2$. 
A few cases of Dynkin diagrams are given in Fig.\ref{An1DynkinTT}.
These can be explicitly written as
\bea
&&\epsilon_a(\theta)=\delta_{a0}\frac{r}{2}e^{\theta}
+\delta_{a,p-2}\frac{r}{2}e^{-\theta}-
\varphi\star\left[\ln\left(1+e^{-\epsilon_{a-1}}\right)+\ln\left(1+e^{-\epsilon_{a+1}}\right)\right](\theta),\nonumber\\
&&{\rm with}\quad\epsilon_a\equiv\epsilon_{a+p-1},\quad{\rm for}\quad a=0,\cdots,p-2.
\label{TTflowTBA}
\eea
Thanks to this additional link, the UV limit $r\to 0$ will be very different from 
\eqref{massflowTBA}. 
We will show in the next section that the UV CFT is the parafermionic Liouville field theory (PLFT) and the $S$-matrices \eqref{masslessRRTT}, \eqref{masslessRLTT}, and \eqref{RLcdd} are those of the parafermionic sinh-Gordon (PShG) model.

\section{Parafermionic sinh-Gordon model}
Parafermions (PFs) \cite{ZamFat} appear in various two-dimensional QFTs.
As fermions can generate supersymmetry,  PFs can generate fractional supersymmetry \cite{ABL}.
In this section, we will focus on the PShG model which generalizes the ordinary sinh-Gordon and the supersymmetric (fermionic) sinh-Gordon (SShG) models.

\def\pd{\partial}
\subsection{The sinh-Gordon model}
This simple integrable QFT has several interesting properties which are shared by both the SShG and PShG models.
The sinh-Gordon model is an integrable QFT with a Lagrangian
\beq
{\cal L}_{\rm ShG}(\Phi)={1\over{4\pi}}(\pd_{a}\phi)^2
+2\mu\,\cosh{2b\phi}.
\label{shGLag}
\eeq
This model can be viewed as a perturbed Liouville field theory (LFT)
\beq
{\cal L}_{L}(\Phi)={1\over{4\pi}}(\pd_{a}\phi)^2
+\mu\,e^{2b\phi},
\label{lftLag}
\eeq
by a relevant operator $e^{-2b\phi}$ where $b$ is the coupling constant and $\mu$ is a dimensionful parameter known as the cosmological constant.
The LFT is a CFT with a central charge
\beq
c_L=1+6Q^2,\qquad Q=b+\frac{1}{b},
\eeq
where $Q$ is a background charge.
The vertex operators 
\beq
V_{\alpha}(x)=e^{2\alpha\phi(x)}
\label{vertex}
\eeq
have conformal dimensions
\beq
\Delta_{\alpha}=\alpha(Q-\alpha).
\label{dimlft}
\eeq
If $\alpha=b$, the vertex operator $e^{2b\phi}$ has the holomorphic dimension $1$ and becomes a screening operator. 

Primary fields are given by \eqref{vertex} with $\alpha=\frac{Q}{2}+iP$ with arbitrary real $P$.
Since the dimension becomes $Q^2/4+P^2$, the two operators with $\alpha=\frac{Q}{2}\pm iP$ have the same dimension and can be identified up to a proportional constant,
namely, 
\beq
V_{Q/2-iP}=R_{L}(P)V_{Q/2+iP}
\eeq
 where $R_L(P)$ is known as the reflection amplitude which can be calculated from the two-point function 
\beq
R_L(P)=(\pi\mu\gamma(b^2))^{-2iP/b}\frac{\Gamma(1+2iPb)\Gamma(1+2iP/b)}{\Gamma(1-2iPb)\Gamma(1-2iP/b)},
\label{reflft}
\eeq
with $\gamma(x)\equiv \Gamma(x)/\Gamma(1-x)$.

The sinh-Gordon model can be described by an exact $S$-matrix between two scalar particles created by the field $\phi(x)$ 
\beq
S_{\rm ShG}(\theta)=\frac{\sinh\theta-i\sin\pi p}{\sinh\theta+i\sin\pi p},\qquad
p=\frac{b}{Q}=\frac{b^2}{1+b^2}.
\label{Sshg}
\eeq
The TBA of the sinh-Gordon model can be derived simply from this $S$-matrix 
\beq
\epsilon(\theta)=mR\cosh\theta-\varphi_{\rm ShG}\star\ln\left(1+e^{-\epsilon}\right)(\theta)
\eeq
where the kernel is a logarithmic derivative of $S_{\rm ShG}$
\beq
\varphi_{\rm shG}(\theta)=\frac{1}{\cosh\left(\theta-i\pi(p-\frac{1}{2})\right)}+\frac{1}{\cosh\left(\theta+i\pi(p-\frac{1}{2})\right)}.
\label{shgkernel}
\eeq

As we will explain in the next section, both the reflection amplitudes and the TBA can be used to derive the scaling functions independently and can be shown to be identical. 
For this comparison, a relation between 
the mass $m$ and the cosmological constant $\mu$,  known as the mass gap relation, is needed  \cite{Alyoshamassmu}
\beq
-\frac{\pi\mu}{\gamma(-b^2)}=\left[\frac{m}{4\sqrt{\pi}}\Gamma\left(\frac{1}{2(1+b^2)}\right)\Gamma\left(1+\frac{b^2}{2(1+b^2)}\right)\right]^{2+2b^2}.
\label{shgmass}
\eeq
Another relevant quantity is the bulk vacuum energy \cite{vacshg} given by
\beq
\mathcal{E}=\frac{m^2}{8\sinh\pi p}.
\label{vacshg}
\eeq
The reflection amplitude, the mass gap relation, and the vacuum energy can be used to compute the ground-state energy $E_0(R)$ independently from the TBA.
This computation will provide cross-checks as we will explain later. 

\subsection{The SShG model}
The $N=1$ supersymmetry maintains the integrability structure of the sinh-Gordon model. 
The supersymmetric Liouville field theory (SLFT) is given by
\beq
{\cal L}_{SL}(\Phi)={1\over{4\pi}}(\pd_{\mu}\phi)^2-\frac{1}{\pi}(\psi{\bar\pd}\psi+{\bar\psi}\pd{\bar\psi})
+4i\mu b^2\,\psi{\bar\psi}\,e^{2b\phi}+\pi\mu^2 b^2\,e^{4b\phi}.
\label{slftLag}
\eeq
The SLFT is a CFT with a central charge
\beq
c_{SL}=\frac{3}{2}+6Q^2,\qquad Q=b+\frac{1}{2b},
\eeq
and the primary fields and their dimensions in the NS sector are given by
\beq
V^{\rm NS}_{P}(x)=e^{2\alpha\phi(x)},\quad V^{\rm R}_{P}(x)=\sigma\,e^{2\alpha\phi(x)},\qquad\alpha=\frac{Q}{2}+iP
\eeq
which have conformal dimensions
\beq
\Delta^{\rm NS}_{\alpha}=\alpha(Q-\alpha)=\frac{Q^2}{4}+P^2,
\label{dimslft}
\eeq
and an additional $1/16$ for the R sector due to the twist field $\sigma$. 
The two operators $V_{\pm P}$ in both sectors are related by the reflection amplitudes \cite{Poghos, RasSta}
\beaq
R^{\rm NS}_{SL}(P)&=&\left(\frac{\pi\mu}{2}\gamma\left(Qb\right)\right)^{-2iP/b}\frac{\Gamma(1+2iPb)\Gamma(1+iP/b)}{\Gamma(1-2iPb)\Gamma(1-iP/b)},\\
R^{\rm R}_{SL}(P)&=&\left(\frac{\pi\mu}{2}\gamma\left(Qb\right)\right)^{-2iP/b}\frac{\Gamma(\frac{1}{2}+2iPb)\Gamma(\frac{1}{2}+iP/b)}{\Gamma(\frac{1}{2}-2iPb)\Gamma(\frac{1}{2}-iP/b)}.
\label{refslft}
\eeaq

The SShG model can be constructed in terms of the $N=1$ super-field $\Phi=\phi+\theta\psi+{\bar\theta}{\bar\psi}+\theta{\bar\theta}F$ with a super-potential $W(\Phi)$ which yields the Lagrangian
\beq
{\cal L}(\Phi)={1\over{4\pi}}(\pd_{\mu}\phi)^2
-{1\over{\pi}}(\psi{\bar\pd}\psi+{\bar\psi}\pd{\bar\psi})
-{i\over{2\pi}}\psi{\bar\psi} W''(\phi)
+{1\over{4\pi}}\left[W'(\phi)\right]^2.
\label{sshgLag}
\eeq
If the superpotential is given by
\beq
W(\phi)=-4\pi\mu\cosh(2b\phi),
\label{massiveSP}
\eeq
the potential energy is $(W')^2\propto \sinh^2(2b\phi)$ which vanishes at $\phi=0$. 
Therefore, the supersymmetry is exact and the on-shell massive particles respect the on-shell supersymmetry, which can be used to find the exact $S$-matrix of the model \cite{ShaWit, Ahnssg}.

However, another SShG model can be defined by
a slightly different super-potential\cite{AKRZ}
\beq
W(\phi)=-4\pi\mu\sinh(2b\phi).
\label{masslessSP}
\eeq
This model shows dramatically different behaviour.
Since $(W')^2\propto \cosh^2(2b\phi)>0$, the supersymmetry is spontaneously broken. The Goldstino,  a  massless fermion, survives stably in all scale while the bosonic particle created by $\phi$ becomes unstable and decays into the chiral (R- and L-moving) fermions as one can see from the
$\psi{\bar\psi} \sinh(2b\phi)$ term in \eqref{RLpart}. We will focus on this massless SShG model.

Since there is no interactions between $R$-fermions (and $L$'s), the $S$-matrix between LL or RR fermions are trivially
\beq
S_{\rm SShG}^{LL}(\th)=S_{\rm SShG}^{RR}(\th)=-1.
\eeq
However, the $S^{LR}$-matrix between the ${L}$- and ${R}$-fermions is non-trivial.
This can be determined by the crossing-unitarity relation
\beq
S_{\rm SShG}^{LR}(\theta)S_{\rm SShG}^{LR}(\theta+i\pi)=1,
\eeq
from which it is derived as
\beq
S_{\rm SShG}^{LR}(\theta)={\sinh\theta-i\sin\pi p\over{\sinh\theta+i\sin\pi p}}.
\label{SRL}
\eeq
The constant $p$ is related to the coupling constant $b$ by
\beq
p=\frac{b}{Q}={2b^2\over{1+2b^2}},
\label{param}
\eeq
with which the mass gap relation is expressed as
\beq
\pi \mu b^2\gamma(bQ)=\left(\frac{m}{8}\frac{\pi p}{\sin\pi p}\right)^{1+2b^2}.
\label{sshgmass}
\eeq
The vacuum energy is given by the same result as \eqref{vacshg}.

This massless scattering theory describes the  $p=3$ (the Ising model) case of the RSOS models,  where $S_{p=3}=-1$ being the  scattering between non-interacting fermions and 
the $S^{LR}$ is nothing but the CDD factor from the $\TTbar$ deformation in \eqref{RLcdd}.
The kernel connecting the $R$ and $L$ nodes in the TBA is again given by \eqref{shgkernel} where $p$ is given by \eqref{param}. 
Since it is shifted by imaginary constants, it is not compatible with the real shift $\alpha$ in \eqref{shiftedkernel} in general. 
Only exception arises when both shifts vanish at self-dual point.
This corresponds to $\alpha=0$ where the kernels connecting the two nodes of $\widehat{A}_2$ are identical 
in Fig.\ref{An1DynkinTT}.

\subsection{Parafermionic Sinh-Gordon models}

Our main claim of this paper is that the $S$-matrices \eqref{masslessRRTT}, \eqref{masslessRLTT}, and \eqref{RLcdd} describe the parafermionic sinh-Gordon model for a generic integer $k$.
So we will define the theory in details.
The Lagrangian of the PLFT can be written as \cite{BasFat}
\beq
{\cal L}_{PL}={\cal L}_{PF}+{1\over{4\pi}}(\pd_{\mu}\phi)^2-\mu
\psi_1{\overline\psi}_1 e^{2b\phi}
+\cdots,
\label{plftLag}
\eeq
where $\psi_1$ and $\overline{\psi}_1$ are $\mathbb{Z}_{k}$ parafermions with (anti-)holomorphic dimension $1-1/k$ and ${\cal L}_{PF}$ denotes the (formal) Lagrangian of the parafermionic CFT.
The ellipsis includes counter terms whose exact expressions are not important in our study.
This PLFT is also a CFT with a central charge
\beq
c_{PL}=\frac{3k}{k+2}+6Q^2,\qquad Q=b+\frac{1}{kb}.
\eeq

The primary fields are given by
\beq
V_P^{(n)}(x)=\sigma_n\, e^{2\alpha\phi},\qquad \alpha=\frac{Q}{2}+iP,\quad n=0,\cdots,k-1
\label{primPL}
\eeq
with the ``spin field'' $\sigma_n$.
The dimension of this is given by
\beq
\Delta_{n}=\frac{Q^2}{4}+P^2+\frac{n(k-n)}{2k(k+2)}
\eeq
where the last term comes from the spin field.
For the case of the SLFT ($k=2$), two cases of $n=0,1$ correspond to the NS and R sectors, respectively.

The effective central charge is defined by $c_{\rm eff}=c-24\Delta_0$ with $\Delta_0$ as a smallest dimension of the theory. 
For the PLFT, this is obtained by $n=0$ and $P=0$,
\beq
c_{\rm eff}=\frac{3k}{k+2}.
\eeq 

For a generic integer $k$, we will focus on the $n=0$ sector which is a generalization of the NS sector for the fractional supersymmetry where the dimension is still the same as \eqref{dimslft}.
Two operators $V^{(0)}_{\pm P}$ are related by the reflection amplitude given by \cite{BasFat}
\begin{equation}
R_{b}^{(k)}(P)=e^{i\delta^{(k)}(P)}=\left(\frac{\pi\mu\gamma(b^{2}+\frac{1}{k})}{k(k^{2}b^{4})^{1/k}}\right)^{-\frac{2iP}{b}}\frac{\Gamma(1+2iPb)\Gamma(1+\frac{2iP}{kb})}{\Gamma(1-2iPb)\Gamma(1-\frac{2iP}{kb})}.
\label{refplft}
\end{equation}
One can check that this reproduces \eqref{reflft} and \eqref{refslft} for $k=1,2$.

Now we can define the PShG models by adding integrable deformations to the PLFT
\beq
{\cal L}_{PL}={\cal L}_{PF}+{1\over{4\pi}}(\pd_{\mu}\phi)^2-\mu
\left(\psi_1{\overline\psi}_1 e^{2b\phi}
+\eta\,\psi_1^{\dagger}{\overline\psi}_1^{\dagger}e^{-2b\phi}\right)
+\cdots,
\label{pshgLag}
\eeq
with the ellipsis including the counter terms.
These models also have both massive and massless phases in the same way as the SShG model.
The PShG model can be considered as the fractional sine-Gordon model, analysed in \cite{ABL}, with an imaginary coupling constant.
Therefore, it has a fractional supersymmetry, which is a generalization of the supersymmetry.
For the massive phase with $\eta=+$, the fractional supersymmetry is maintained, and the $S$-matrix describes scatterings among massive multiplets. 
On the other hand, if it is massless with $\eta=-$, the fractional supersymmetry is broken spontaneously and only massless PFs will be left to describe the theory.

The mass gap relation has been also computed in \cite{BasFat}:
\beq
\frac{\pi\kappa}{k}\gamma(bQ)=\left[\frac{m}{8\Gamma(\frac{k+2}{2})} \Gamma\left(1+\frac{k^2b^2}{2(1+kb^2)}\right)\Gamma\left(\frac{k}{2(1+kb^2)}\right)\right]^{2bQ},
\label{massgap}
\eeq
and the vacuum energy is the same as \eqref{vacshg} with $p=b/Q$.

We claim that the $S$-matrices between these massless particles are given by the RSOS $S$-matrices \eqref{masslessRR} and \eqref{RLcdd} with the self-dual coupling constant $b=1/\sqrt{k}$.
This will be be justified by comparing scaling functions from the TBA  with those obtained by the reflection amplitudes.

\section{Comparing the TBA to the reflection amplitudes}

In this section we compare the finite size ground-state energy coming from the reflection amplitudes to the same quantity coming from the TBA.  

\subsection{Effective central charge from the reflection amplitudes}

In the parefermionic sinh-Gordon model the effective central charge is governed by the primary field with $n=0$ and with the minimum value of the momentum $P$ 
\begin{equation}
c_{\rm eff}(R)={3k\over{k+2}}-24P^2+{\cal O}(R).
\end{equation}
Since the primary field is confined in Liouville potentials on both sides, the momentum is quantized by the condition
\begin{equation}
\delta^{(k)}(P)=\delta_{1}P+\delta_{3}P^{3}+\dots=\pi+4QP\ln x,\qquad x=\frac{R}{2\pi},
\end{equation}
where the reflection phase $\delta^{(k)}$ is defined in \eqref{refplft}.
At small volume the momentum can be expanded as  
\begin{equation}
P=-\frac{\pi}{4Q\ln x}-\frac{\pi\delta_{1}}{16Q^{2}\ln^{2}x}-\frac{\pi\delta_{1}^{2}}{64Q^{3}\ln^{3}x}-\frac{\pi\delta_{1}^{3}+\pi^{3}\delta_{3}}{256Q^{4}\ln^{4}x}+\dots
\end{equation}
The corresponding effective central charge has a logarithmic volume dependence
\begin{equation}
c_{\mathrm{eff}}=\frac{3k}{k+2}-\frac{3\pi^{2}}{2Q^{2}\ln^{2}x}-\frac{3\pi^{2}\delta_{1}}{4Q^{3}\ln^{3}x}-\frac{9\pi^{2}\delta_{1}^{2}}{32Q^{4}\ln^{4}x}-\frac{3(2\pi^{2}\delta_{1}^{3}+\pi^{4}\delta_{3})}{64Q^{5}\ln^{5}x}+\dots\label{eq:ceff}
\end{equation}
Observe that the $(\ln x)^{-1}$ term is missing. 

The  small $P$ expansion takes the form
\begin{equation}
\delta^{(k)}(P)=-\frac{2P}{b}\ln\frac{\pi\mu\gamma(b^{2}+\frac{1}{k})}{k(k^{2}b^{4})^{1/k}}-4PQ\gamma_{E}+P^{3}\zeta(3)\frac{16}{3}(b^{3}+\frac{1}{b^{3}k^{3}})+\dots
\end{equation}
In order to get the final form we still need to use the massgap relation \eqref{massgap}. 
Let us note that the mass appears only at the linear order as
\begin{equation}
\delta^{(k)}(P)=-4PQ\ln m+\dots
\end{equation}
which nicely combines with $R$ to the dimensionless volume $mR=r$. 

In order to compare with the TBA analysis we specify the results for the self dual point, defined by
\begin{equation}
b^{2}=\frac{1}{k}\quad;\quad Q=\frac{2}{\sqrt{k}}.
\end{equation}
The mass gap relation simplifies considerably 
\begin{equation}
\frac{\pi\mu}{k}\gamma(2/k)=\left[\frac{m\Gamma^{2}(\frac{k}{4})}{16\Gamma(\frac{k}{2})}\right]^{4/k}
\end{equation}
and the reflection factor 
\begin{equation}
e^{i\delta(P)}=\left(\frac{m\Gamma^{2}(\frac{k}{4})}{16\Gamma(\frac{k}{2})}\right)^{-\frac{8iP}{\sqrt{k}}}\frac{\Gamma(1+\frac{2iP}{\sqrt{k}})^{2}}{\Gamma(1-\frac{2iP}{\sqrt{k}})^{2}}.
\end{equation}
The small $P$ expansion of the phase at the self dual point is
\begin{equation}
\delta^{(k)}(P)=\delta_{1}P+\delta_{3}P^{3}+\dots=-\frac{8}{\sqrt{k}}\ln\left(\frac{m\Gamma^{2}(\frac{k}{4})}{16\Gamma(\frac{k}{2})}\right)P-8\frac{\gamma_{E}}{\sqrt{k}}P+\frac{32\zeta(3)}{3k^{\frac{3}{2}}}P^{3}+\dots.
\end{equation}
Plugging back these values into eq. \eqref{eq:ceff} and taking into
account that $x=R/2\pi$ leads to the coefficients in Table \ref{ccoeffs}. 
We compare these numbers with the same quantities obtained by numerically solving the TBA equations.

\subsection{Numerical solution of the TBA }

\begin{figure}
\begin{centering}
\includegraphics[width=8cm]{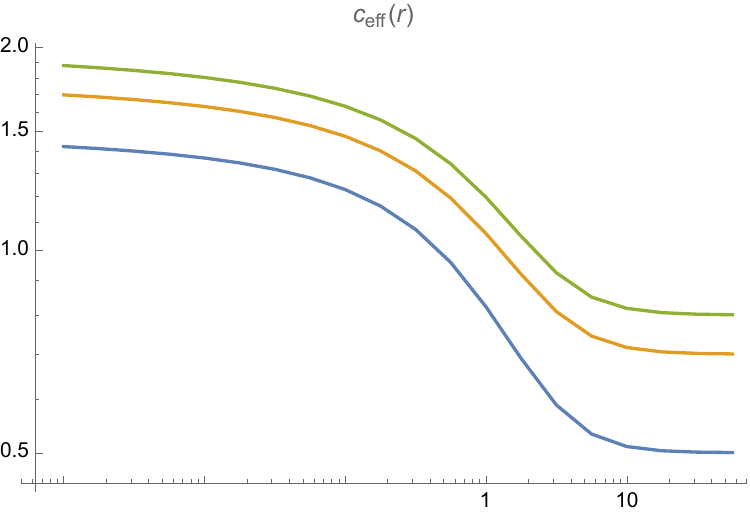}
\par\end{centering}
\caption{Logarithmic plot of the effective central charge as the function of
the dimensionless volume $r=\widehat{M}R$ for $k=2,3,4$ (blue,orange,green).}

\label{ceffr}
\end{figure}

We solve the TBA equations \eqref{TTflowTBA}, by discretizing the pseudo energies
$\epsilon_{a}$ and performing the convolutions using discrete Fourier
transforms. Iterating the equations until reaching the prescribed
precision, (which we choose to be $10^{-16}$), leads to the effective
central charge $c_{\mathrm{eff}}(r)$ as the function of the dimensionless
volume $r=\widehat{M}R$, which serves as the RG parameter. We present
the results for the $k=p-1=2,3,4$ cases. The behaviour of the effective central charge is displayed on Fig.
\ref{ceffr}.

In the IR, i.e. for large volumes $R$, the left and right movers
decouple and we get back the IR minimal model CFT with central charge
$c_{\mathrm{IR}}=1-\frac{6}{(k+1)(k+2)}$. By investigating the numerical solution one can observe
that the non-trivial behaviour is concentrated on the small and large
$\theta$ region. In each domain one of the colored nodes (with the
driving term $\epsilon_{0}$ or $\epsilon_{p-2}$ ) becomes negligible.
The TBA is then no longer the ring, rather the line, which describes
the same left or right moving scattering theories \eqref{CFTTBA}. 
\begin{figure}
\begin{centering}
\includegraphics[width=8cm]{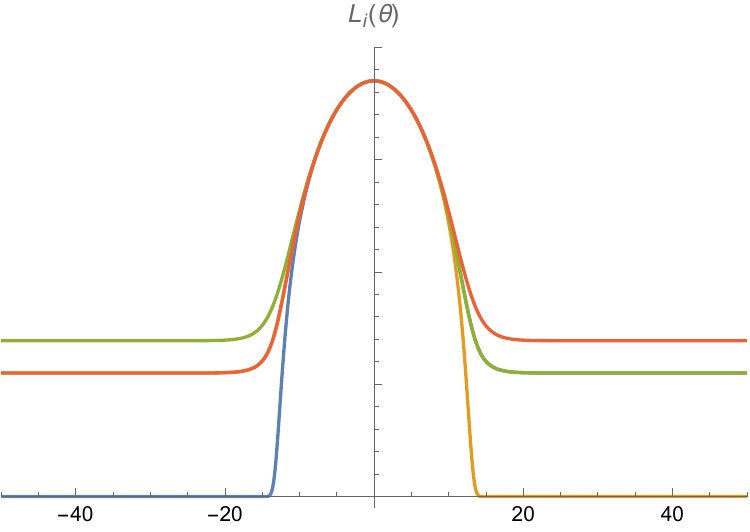}
\par\end{centering}
\caption{Plot of the $L_{i}=\log(1+e^{-\epsilon_{i}})$ functions for $k=4$
and dimensionless volume $r=10^{-5}$ as the function of $\theta$.
The functions $\{L_{0},L_{1},L_{2},L_{3}\}$ correspond to \{blue,orange,green,red).
Clearly $L_{i}(\theta)=L_{3-i}(-\theta)$.}

\label{LiUV}
\end{figure}

In the UV, the effective central charge approaches its UV value $c_{\mathrm{UV}}=\frac{3k}{k+2}$
very slowly. The reason is that in the central domain the functions
$L_{i}(\theta)=\log(1+e^{-\epsilon_{i}(\theta)})$ does not approach
any plateaux value, see Fig.\ref{LiUV}. Indeed in this limit the
driving terms are negligible and the central behaviour is governed
by the same function we would have in  the sinh-Gordon theory at the self dual point, where 
it is well-known that the effective central charge approaches
its UV value logarithmically. We have already obtained this logarithmic behaviour in our case, which
we try to fit numerically now. 
We  thus parametrize the small volume behaviour of the central charge as 
\beq
c_{{\rm eff}}(r)=\frac{3k}{k+2}+\sum_{n=2}\frac{c_{n}(k)}{(\log r)^{n}}+O(r)
\eeq
by focusing on the logarithmic corrections and neglecting any higher
order polynomials in $r$. We extract the coefficients $c_{n}(k)$
numerically, by fitting $c_{\mathrm{eff}}(r)$ in the range $10^{-10}-10^{-6}.$
The results are displayed in the Table \ref{ccoeffs} and shows a convincing agreement
with the analytically obtained expressions from the reflection factors. 

\begin{table}
\begin{centering}
\begin{tabular}{|c|c|c|c|c|}
\hline 
 & $c_{2}$ & $c_{3}$ & $c_{4}$ & $c_{5}$\tabularnewline
\hline 
\hline 
$k=2$ & $7.402199$ & $42.7620$ & $185.218$ & $714.563$\tabularnewline
\hline 
 & $7.402203$ & $42.7628$ & $185.282$ & $717.247$\tabularnewline
\hline 
$k=3$ & $11.10332$ & $77.8573$ & $409.598$ & $1924.84$\tabularnewline
\hline 
 & $11.10330$ & $77.8543$ & $409.425$ & $1919.36$\tabularnewline
\hline 
$k=4$ & $14.8045$ & $119.428$ & $722.797$ & $3898.75$\tabularnewline
\hline 
 & $14.8044$ & $119.4197$ & $722.475$ & $3892.55$\tabularnewline
\hline 
\end{tabular}
\par\end{centering}
\caption{Numerically fitted coefficients in the various cases above and their
analytical expressions from the reflection amplitudes, below. }

\label{ccoeffs}
\end{table}

The agreement we found tests not only the approach based on the reflections factors
of the PShG model, but specifically its mass gap relation and the
first two coefficients $\delta_{1},\delta_{3}$. 

\section{Roaming TBA}
\begin{figure}
\begin{centering}
\includegraphics[width=13cm]{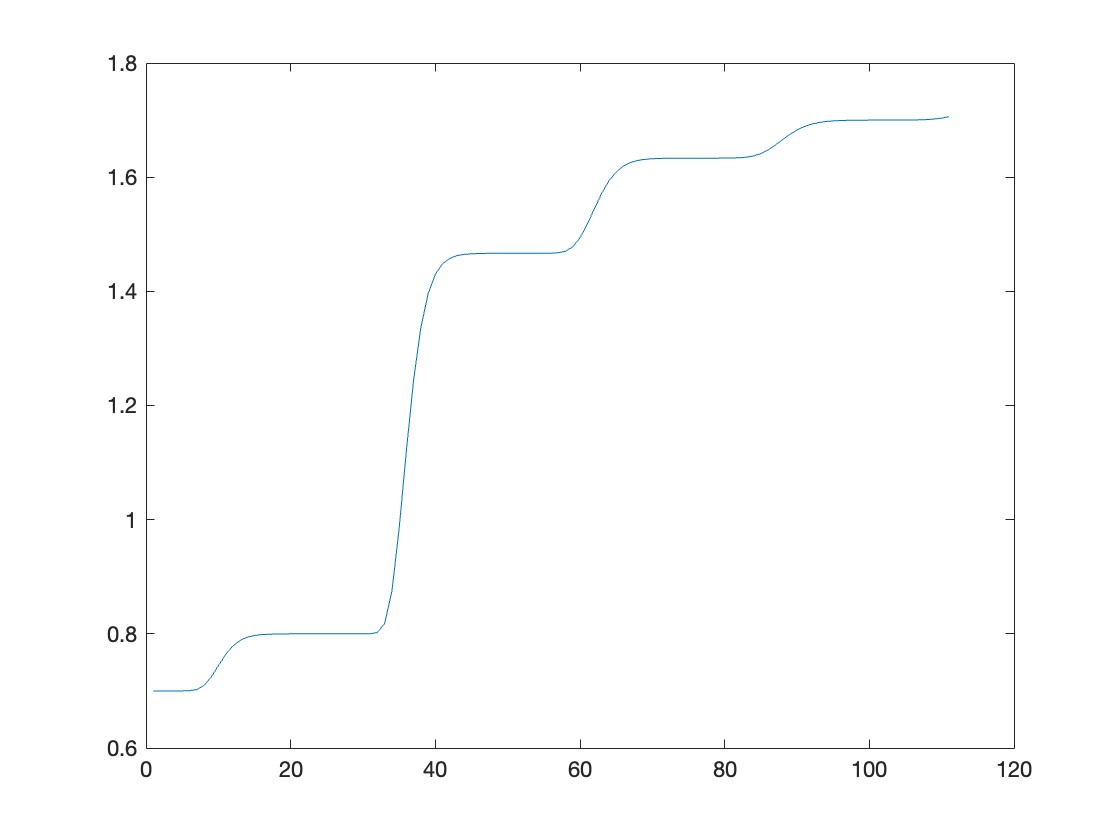}
\par\end{centering}
\caption{$c_{\rm eff}(r)$ vs $-\log r $ for $p=4$ with $\alpha=10\pi$.}
\label{roamp4}
\end{figure}

So far we have considered $S^{RL}$ and $S^{LR}$ matrices given by \eqref{RLcdd} with $\alpha=0$.
For the case of $\alpha\neq 0$, the $S^{RL}$ and $S^{LR}$ are changed by a shift $\alpha$ in the arguments.
The TBA can be derived from the same set of massless $S$-matrices as
\bea
\epsilon_a(\theta)&=&-
\varphi\star\left[\ln\left(1+e^{-\epsilon_{a-1}}\right)+\ln\left(1+e^{-\epsilon_{a+1}}\right)\right](\theta),
\quad a=1,\cdots,p-3,\\
\epsilon_0(\theta)&=&\frac{r}{2}e^{\theta}-
\varphi\star\ln\left(1+e^{-\epsilon_{1}}\right)+\varphi^{(+)}\star\ln\left(1+e^{-\epsilon_{p-2}}\right)(\theta),\\
\epsilon_{p-2}(\theta)&=&\frac{r}{2}e^{-\theta}-
\varphi\star\ln\left(1+e^{-\epsilon_{p-3}}\right)+\varphi^{(-)}\star\ln\left(1+e^{-\epsilon_{1}}\right)(\theta),
\label{TTroamingTBA}
\eea
with kernels $\varphi^{(\pm)}$ defined by
\beq
\varphi^{(\pm)}(\theta)=\varphi(\theta\pm\alpha)
\eeq
and with the same graph in Fig.\ref{An1DynkinTT}.

For small $\alpha$, the TBA shows qualitatively same behaviour as $\alpha=0$, namely, interpolating the minimal CFT $\mathcal{M}_p$ with the $\mathbb{Z}_{p-1}$ PShG model.
However, for sufficiently large $\alpha\gg 1$, this TBA is showing an interesting roaming trajectories ($p=k+1$):
\beq
\mathcal{M}_{k+1}\to \mathbb{Z}_{k}\mathcal{M}_1\to \mathbb{Z}_{k}\mathcal{M}_{k+1}\to \mathbb{Z}_{k}\mathcal{M}_{2k+1}
\to \cdots,
\eeq
where we denote $\mathbb{Z}_{k}$ PF minimal series by $\mathbb{Z}_{k}\mathcal{M}_{\ell}$ ($\ell=1,2,\cdots$) which can be written as coset CFTs as follows:
\beq
\mathbb{Z}_{k}\mathcal{M}_{\ell}=\frac{su(2)_k\otimes su(2)_{\ell}}{su(2)_{k+\ell}},\qquad
{\rm with}\quad c=\frac{3k\ell(k+\ell+4)}{(k+2)(\ell+2)(k+\ell+2)}.
\eeq
For example, as shown in Fig.\ref{roamp4} for $p=4$ case, the starting IR CFT has $c_{IR}=0.7$ of the $\mathcal{M}_4$ CFT.
The next central charge jumps to $c=\frac{4}{5}$ which is the first CFT in $\mathbb{Z}_3$ PF minimal series and 
succeeded by  $c=\frac{22}{15},\frac{49}{30},\cdots$.

This TBA is part of the TBA systems conjectured in \cite{DorRav} to describe roaming trajectories between coset minimal models. Although those kernels are apparently different, they can be transformed to those in 
\eqref{TTroamingTBA} by shifting the rapidities in the definition of the pseudo energies and by redefining the scale $r$ appropriately.
Therefore, the massless $S$-matrices in 
\eqref{masslessRRTT}, \eqref{masslessRLTT}, and \eqref{RLcdd} are the exact $S$-matrices behind  the 
conjectured roaming TBA of $\mathbb{Z}_{k}$ PF minimal series.

\section{Conclusion}

QFTs that interpolate between two CFTs in their UR and IR limits are valuable but rare examples from which we can understand quantitatively how fundamental degrees of freedom such as operators and on-shell particles are connected.
In this paper, we have approached to find new QFTs from  the IR point of view based on the exact massless $S$-matrices which are deformed by $\TTbar$. 
These special irrelevant fields preserve the integrability and modify the $S$-matrices between $R$- and $L$-particles in a systematic way. 
Generalizing \cite{AhnLeC}, we have applied $\TTbar$ deformations on non-diagonal kink scattering theories of the perturbed minimal CFTs $\mathcal{M}_p$.
We have found that only two fine-tuned $\TTbar$ deformations can be UV complete. 
The first one is the $\mathbb{Z}_{p-1}$ PShG model with the self-dual coupling constant which leads to the $\mathbb{Z}_{p-1}$ PF LFTs in the UV limit.
Another one is covering certain $\mathbb{Z}_{p-1}$ parafermionic minimal CFT series with roaming trajectories.
It is remarkable to see how a fractional supersymmetry associated with the $\mathbb{Z}_{p-1}$ PF emerges from the  simple minimal CFT $\mathcal{M}_p$ by the fined tuned irrelevant $\TTbar$ deformations.
This emergent symmetry generalizes the phenomena observed in the Ising model ($p=3$) \cite{AKRZ}.

We want to emphasize that our TBAs have been derived from exact massless $S$-matrices rather than many educated guesses on TBAs and non-linear integral equations in the literature (See \cite{Dorey, Clare} and references therein.)
In this work, we have derived two of previously conjectured TBAs, one in Fig.\ref{D_type} and the PF roaming TBA, from the exact $S$-matrices.
It would be nice if we can prove other conjectured TBAs in this way.

In this work, we have considered the $\mathcal{M}_p$ CFT as a scattering theory of RSOS kinks based on the $\phi_{1,3}$ deformation.
In fact, there are other integrable descriptions of the same minimal CFTs related to different integrable deformations. 
It would be interesting to find new UV CFTs based on these different $S$-matrices associated with the same IR minimal CFTs. In this way, we may lead to a complete classification of UV complete theories for a given IR CFT.

Recently massless scattering theories gain attentions related to the world-sheet $S$-matrices of AdS3/CFT2 duality
\cite{adscft}.
Being CFTs, these $S$-matrices are between $RR$ and $LL$ particles while $RL$ scatterings are trivial.
It will be interesting to consider non-trivial $RL$ scatterings, possibly related to the $\TTbar$ deformations and their RG flows in the context of AdS/CFT duality.
Another interesting direction is to understand relations between these new RG flows and non-invertible symmetries associated with some topological defect lines \cite{CLSWY}.

\section*{\bf Acknowledgement}
We want to thank J. Balog, P. Dorey, M. Lencses, F. Ravanini for valuable discussions and comments.
CA thanks the mathematical physics group of Matthias Staudacher at Humboldt University in Berlin and Wigner Institute in Budapest and ZB thanks Ewha University for hospitality where parts of this work have been performed. 
In particular, CA acknowledges partial support of his stay by the {\it Kolleg Mathematik Physik Berlin} (KMPB).
This work was supported in part by the government of the Republic of Korea (MSIT) and the 
National Research Foundation of Korea  (NRF-2023K2A9A1A01098567) for the Mobility program between 
Korea and Hungary, by (NRF-2016R1D1A1B02007258) (CA), and by the K134946 NKFIH Grant.


\begin{thebibliography}{99}

\bibitem{sasha} A. B. Zamolodchikov, Int. J. Mod. Phys. {\bf A4} (1989) 
4235.
\bibitem{zamolRG}  A. B. Zamolodchikov, 
{\em Irreversibility of the flux of the renormalization group in a 2D field theory},
Pis'ma Eksp. Teor. Fiz. {\bf 43} (1986) 565 
\bibitem{alyosha_tba} Al. B. Zamolodchikov,  {\em Thermodynamic Bethe ansatz in Relativistic Models: Scaling 3-state Potts and Lee-Yang Models}, Nucl. Phys. {\bf B342} (1990) 695
\bibitem{alyosha_tba_flow}Al.B. Zamolodchikov, {\em From Tricritical Ising to Critical Ising By Thermodynamic Bethe ansatz}, Nucl. Phys. {\bf B358} (1991) 524
\bibitem{alyosha_tba_cosetflow}Al.B. Zamolodchikov, {\em TBA Equations for Integrable Perturbed $SU(2)_k\times  SU(2)_l/SU(2)_{k+l}$ Coset Models}, Nucl. Phys. {\bf B366} (1991) 122
\bibitem{zamzam_flow}A.B. Zamolodchikov and Al.B. Zamolodchikov, {\em Massless factorized scattering and sigma models with topological terms}, Nucl. Phys. {\bf B379} (1992) 602
\bibitem{FSZ} P. Fendley, H. Saleur, and Al. B. Zamolodchikov, {\em Massless Flows II: the exact S-matrix approach}, Int. J. Mod. Phys. {\bf A8} (1993) 5751, arXiv:hep-th/930405.
\bibitem{Fateev} V. A. Fateev, {\em Integrable perturbations of $Z_N$ parafermion models and the $O(3)$ sigma model}, Phys. Lett. {\bf B271} (1991) 91.

\bibitem{DDT}P. Dorey, C. Dunning and R. Tateo, {\em New families of flows between two-dimensional conformal field theories}, Nucl. Phys. {\bf B578} (2000) 699
\bibitem{AKRZ}C. Ahn, C. Kim, C. Rim, and Al.B. Zamolodchikov, {\em RG flows from super-Liouville theory to critical Ising model}, Phys. Lett. {\bf B541} (2002) 194
\bibitem{AhnLeC} C. Ahn and A. LeClair, {\em On the classification of UV completions of integrable $ T{\bar T}$  deformations of CFT}, JHEP {\bf 2022} (2022) 179, arXiv:2205.10905 [hep-th]
\bibitem{SZ} 
  F.A. Smirnov and A.B. Zamolodchikov,  {\em On the space of integrable quantum field theories}, Nucl. Phys. {\bf  B915}  (2017) 363, arXiv:1608.05499 [hep-th]
\bibitem{CNST} 
  A. Cavagli\`a, S. Negro, I.M. Szecsenyi and R. Tateo, {\em $T{\bar T}$-deformed 2D quantum field theories}, JHEP {\bf 10}  (2016) 112, arXiv:1608.05534 [hep-th]
\bibitem{BerLeC} D. Bernard and A. LeClair, {\em Residual quantum symmetries of the restricted sine-Gordon theories}, Nucl. Phys. {\bf B340} (1990) 721
\bibitem{Shota} C. Copetti, L. Cordova and S. Komatsu, {\em Non-Invertible Symmetries, Anomalies and Scattering Amplitudes}, arXiv:2403.04835 [hep-th]
\bibitem{alyosha_tba_rsos}Al.B. Zamolodchikov, {\em Thermodynamic Bethe ansatz for RSOS scattering theories}, Nucl. Phys. {\bf B358} (1991) 497
\bibitem{Alyoshamassmu}Al.  B.  Zamolodchikov,  {\em Mass Scale in the sine-Gordon model and its Reductions}, Int. J.  Mod.  Phys.  {\bf A10} (1995) 1125

\bibitem{ZamFat} A.B. Zamolodchikov and V.A. Fateev, Soy. Phys. JETP {\bf 62} (1985) 215

\bibitem{ABL}  C. Ahn, D. Bernard, and A. LeClair, {\em Fractional Supersymmetries in Perturbed coset CFTs
and Integrable Soliton Theory}, Nucl. Phys. {\bf B346} (1990) 409
\bibitem{vacshg} C. Deatri and H. deVega. Nucl.Phys. {\bf B358} (1991) 251
\bibitem{Poghos} R. H. Poghossian, {\em Structure constants
in the N=11 super-Liouville field theory}, Nucl. Phys. {\bf B496} (1997) 451
\bibitem{RasSta} R. C. Rashkov and M. Stanishkov, {\em Three-point correlation functions in N=1 super Liouville theory}, Phys. Lett. {\bf B380} (1996) 49

\bibitem{ShaWit} R. Shankar and E . Witten, Phys . Rev . {\bf D17} (1978) 2134
\bibitem{Ahnssg}  C. Ahn, {\em Complete S-matrices OF Supersymmetric Sine-Gordon Theory
and Perturbed Superconformal Minimal Model}, Nucl. Phys. {\bf B354} (1991) 57
\bibitem{BasFat} P. Baseilhac and V. A. Fateev, {\em Expectation values of local fields for a two-parameter family of integrable models and related perturbed conformal field theories}, Nucl. Phys. {\bf B532} (1998) 567
\bibitem{DorRav}  P. Dorey and F. Ravanini, {\em Generalising the staircase models}, Nucl. Phys. {\bf B406} (1993) 708, arXiv:9211115 [hep-th]; {\em Staircase Models from Affine Toda Field Theory}, Int. J. Mod. Phys. {\bf A8} (1993) 873, arXiv:9206052 [hep-th]


\bibitem{adscft} S. Frolov and  A. Sfondrini, {\em Massless S matrices for AdS3/CFT2}, JHEP {\bf 04} (2022) 067, arXiv:2112.08895 [hep-th]
\bibitem{Dorey} P. Dorey, {\em New families of flows between two-dimensional conformal field theories}, Nucl. Phys. {\bf B578} (2000) 699 arXiv:0001185 [hep-th]
\bibitem{Clare}  C. Dunning, {\em Massless flows between minimal W models}, Phys. Lett. {\bf B537} (2002) 297, arXiv:0204090 [hep-th]
\bibitem{CLSWY} C. Chang, Y. Lin, S. Shao, Y. Wang, and X. Yin,  {\em Topological Defect Lines and Renormalization Group Flows in Two Dimensions}, JHEP {\bf 03}  (2019)  26, arXiv:1802.04445 [hep-th]

\end{thebibliography}
\end{document}